\documentclass[12pt]{iopart}

\usepackage{enumerate}
\usepackage{amssymb}

\usepackage{amsthm}

\newcommand{\be}{\begin{equation}}
\newcommand{\ee}{\end{equation}}
\newcommand{\ba}{\begin{eqnarray}}
\newcommand{\ea}{\end{eqnarray}}

\begin{document}

\title[The classical Kepler problem and geodesic motion on spaces of constant curvature]
{The classical Kepler problem and geodesic motion on spaces of constant curvature}

\author{Aidan J. Keane, Richard K. Barrett and John F.L. Simmons}
\address{
Department of Physics and Astronomy, University of Glasgow,
Glasgow G12 8QQ, Scotland, UK
}

\begin{abstract}
In this paper we clarify and generalise previous work by Moser and Belbruno
concerning the link between the motions in the classical Kepler problem and
geodesic motion on spaces of constant curvature.
Both problems can be formulated as Hamiltonian systems and the phase flow in each
system is characterised by the value of the corresponding Hamiltonian and one other
parameter (the mass parameter in the Kepler problem and the curvature parameter
in the geodesic motion problem). Using a canonical transformation the Hamiltonian vector field for the geodesic motion problem is transformed into one which is proportional to that for the Kepler problem.
Within this framework the energy of the Kepler problem is equal to (minus)
the curvature parameter of the constant curvature space and the mass parameter
is given by the value of the Hamiltonian for the geodesic motion problem.
We work with the corresponding family of evolution spaces and present a unified treatment which is valid for all values of energy continuously. As a result, there is a correspondence between the constants of motion for both systems and the Runge-Lenz vector in the Kepler problem arises in a natural way from the isometries of a space of constant curvature. In addition, the canonical nature of the transformation guarantees that the Poisson bracket Lie algebra of constants of motion for the classical Kepler problem is identical to that associated with geodesic motion on spaces of constant curvature.
\end{abstract}

\pacs{95.10.Ce, 45.50.Pk, 04.20.-q}

\section{Introduction}
It is well known that the classical Kepler problem is an example of a dynamical system with a ``hidden" dynamical symmetry: The system is obviously spherically symmetric but the existence of the
Runge-Lenz vector means the Lie algebra of constants of motion can be extended
to $so(4)$, $iso(3)$ and $so(3,1)$ for negative, zero and positive energy
orbits respectively.
Thus there is a one-to-one correspondence between the dynamical symmetry
algebras of motion in the classical Kepler problem (with Hamiltonian
function $H$) and geodesic motion on three-dimensional spaces of constant
curvature (with Hamiltonian function $G$) for each energy surface (i.e.,
the three-sphere ${\cal S}^3$ with positive curvature, Euclidean three-space
${\cal E}^3$ with zero curvature and the three-hyperboliod ${\cal H}^3$ with
negative curvature).

Moser \cite{moser70} addressed the geometrical nature of the energy surface for the
Kepler problem for negative values of energy and claims that for a negative constant $E$, the energy surface $H=E$ can be mapped topologically one-to-one
into the unit tangent bundle of the n-dimensional sphere ${\cal S}^n$ with the
north pole excluded.
The flow defined by the Kepler problem is mapped into the geodesic flow
on the punctured sphere after a change of independent variable.
This is extended by Belbruno \cite{belbruno77} to include positive energy orbits
$E >0$ in the Kepler problem and the three-hyperboloid ${\cal H}^3$ and also for the case with $E=0$ and
three-dimensional Euclidean space ${\cal E}^3$.
Osipov \cite{osipov77} also tackles the case of positive energies $E>0$.
More recently, Mes\'{o}n and Vericat \cite{meson98} have considered the case of a repulsive field.

In this paper we extend the results of Moser \cite{moser70} and Belbruno \cite{belbruno77}
by introducing a family of evolution spaces rather than considering a single Kepler problem.
We then present a transformation allowing us to relate this space of
{\it all} classical Kepler problems with distinct mass parameters $\alpha$ to
the space of {\it all} Hamiltonian systems of free particle motion on spaces of
constant curvature $k$. Thus the phase flow in each system is characterised by
{\it two} parameters, the constant energy $H=E$ (or $G=C$) and an additional
constant $\alpha$ (or $k$).
We demonstrate (via a convenient choice of canonical coordinates) that the phase
flows for the two systems can be considered parallel.
Within this framework the energy of the Kepler problem is equal to (minus)
the curvature parameter of the constant curvature space and the mass parameter
is given by the value of the Hamiltonian for the geodesic motion problem.
The fact that the Hamiltonian vector fields can be considered to be parallel ensures the correspondence of
the constants of motion for both systems.
This result gives the Runge-Lenz vector a ``purely geometrical"
interpretation in that it arises from isometries of a three-dimensional manifold
associated with the Kepler Hamiltonian system.
In addition, the canonical nature of the transformation guarantees the correspondence between the Poisson
bracket Lie algebras of constants of motion on the constant energy hypersurfaces
of the two systems.
Our analysis is valid for all values of energy $E$ continuously and is thus a unified
treatment which is independent of the sign of the energy.
This is in contrast to the work of Moser \cite{moser70} and Belbruno \cite{belbruno77} which
dealt with a {\it single} Kepler problem and necessitated the use of a different
procedure for each sign of energy.

The dynamical features of the Kepler problem can be analysed using alternative techniques.
Iwai \cite{iwai81} defines a four-dimensional conformal Kepler problem in order to associate the
three-dimensional Kepler problem with the four-dimensional harmonic oscillator.
The Hamiltonian vector fields corresponding to the motion in the four-dimensional conformal Kepler problem
and the four-dimensional harmonic oscillator are shown to be parallel on appropriate energy surfaces and then
the four-dimensional conformal Kepler problem is reduced to the three-dimensional Kepler problem.
Thus the (negative) energy surface in the three-dimensional Kepler problem is obtained together with
the $SO(4)$ symmetry from an appropriate energy surface of the four-dimensional harmonic oscillator.
Mladenov \cite{mladenov89} applies this technique to the MIC-Kepler problem (motion in the dual charged
Coulomb field modified by a centrifugal term) and then applies a geometric quantization scheme to the
extended phase space of the MIC-Kepler problem.
Guillemin and Sternberg \cite{guillemin90} show that the Kepler motion can be enlarged to geodesic flow
on a curved Lorentzian five-dimensional manifold. In their work the mass parameter $\alpha$ is directly
related to a conjugate momentum coordinate in the cotangent bundle.
The advantage of the method described in the present work is that the dynamical features of the
Kepler problem are derived directly from those of a system with the {\it same dimension}.
In addition, the features of the classical Kepler problem have a direct geometrical
origin in that they arise from the
properties of the geodesics of a three-dimensional manifold of constant curvature.

There is an extensive amount of material in the literature relating to the geometry of the Kepler
problem and we refer the reader to Guillemin and Sternberg \cite{guillemin90} and Milnor \cite{milnor83}
for an overview.

We briefly outline some concepts necessary for the description of Hamiltonian
systems in section \ref{sec:hamsystems}.
In section \ref{sec:keppro} we discuss the classical Kepler problem
and its symmetries and constants of motion.
Geodesic motion on constant curvature spaces and canonical transformations of
the associated phase space are considered in section \ref{sec:geomotconcurgeo}.
Then in section \ref{sec:dynsys} we construct a map between these two
dynamical systems and clarify their relationship.

\section{Hamiltonian systems}\label{sec:hamsystems}
A symplectic manifold $N$ endowed with a closed nondegenerate symplectic
two-form ${\bf {\tilde \omega}}$ is denoted $(N, {\bf {\tilde \omega}})$.
Let us introduce the $2n$ coordinates $(x^1, \dots, x^n$, $p_1, \dots, p_n)$.
Then the canonical one-form ${\bf \tilde{p}}$ on the symplectic manifold
$(N, {\bf {\tilde \omega}})$ is defined as
\[
{\bf \tilde{p}} = p_i \; {\bf d}x^i ,
\]
and the canonical symplectic two-form
\[
{\bf d} {\bf \tilde{p}}
= {\bf d} p_i \wedge {\bf d} x^i.
\]
The symplectic two-form ${\bf {\tilde \omega}}$ can always be written locally as
${\bf {\tilde \omega}}={\bf d}{\bf \tilde{p}}$.
For a symplectic manifold $(N,{\bf {\tilde \omega}})$, the Hamiltonian
vector field ${\bf {\hat X}}_f$ corresponding to the function $f$ is defined
as the unique smooth vector field on $N$ satisfying
\[
{\bf {\tilde \omega}} \, ({\bf {\hat X}}_f) = - {\bf d} f.
\]
A Hamiltonian system is a symplectic manifold $(N, {\bf {\tilde \omega}})$
endowed with a Hamiltonian function $H$ and denoted
$(N, {\bf {\tilde \omega}}, H)$.

Consider the direct product space $W= \mathbf{R} \times N$ which is a $2n+1$
dimensional manifold locally described by the coordinates
$(x^1, \dots, x^n$, $p_1, \dots, p_n, \tau)$. Then we can define
a closed two-form on $W$ by
\be
{\bf \Omega}={\bf d}p_i \wedge {\bf d}x^i - {\bf d}H \wedge d\tau,
\ee
where $H$ is a function on $W$. Then $(W, {\bf \Omega})$ is said to be an
evolution space, see Andri{\'{e}} \cite{andrie76} for further details.

\section{The classical Kepler Problem} \label{sec:keppro}
The classical Kepler problem is the motion of a particle in Euclidean
(configuration) space ${\cal E}^3$ under a central inverse square force
$-\alpha / |{\bf x}|^2$.
The singularity $|{\bf x}|=0$ is removed from the manifold ${\cal E}^3$:
The configuration space is taken to be ${\cal E}^3 - \{ 0 \}$.
The corresponding phase space is the cotangent bundle
$({\cal E}^3 - \{ 0 \}) \times \mathbf{R}^3$.
Thus the classical Kepler problem is the Hamiltonian system
$(N,{\bf {\tilde \omega}},H)$ where $N$ is the cotangent bundle
$({\cal E}^3 - \{ 0 \}) \times \mathbf{R}^3$ and $H$ is the Hamiltonian
function
\be
H={|{\bf p}|^2 \over 2} - {\alpha \over {|{\bf x}|}}
\label{eqn:hamckp}
\ee
with $\alpha$ a constant. The Hamiltonian vector field corresponding to
the phase flow is
\be
{\bf {\hat X}}_{H} =
 p_i{{\partial \phantom{z}} \over {\partial x^i}} -
\alpha {x^j \over |{\bf x}|^3} {{\partial \phantom{y}} \over {\partial p_j}},
\label{eqn:hamvecckp}
\ee
where $i,j=1,2,3$.
The system $(N,{\bf {\tilde \omega}},H)$ only shares the rotational
symmetry of the underlying base manifold ${\cal E}^3$, however it does admit an
extra set of constants of motion $A_i$. These quantities are the components of
the {\it Runge-Lenz vector} which determines the orientation of the major axis
in the orbital plane.
The quantities
\ba
J_i = \epsilon_{ij}^k \, x^j \, p_k,
\nonumber\\
\sqrt{2} A_i =
x^i \biggl(|{\bf p}|^2 - {\alpha \over {|{\bf x}|}} \biggr)
-  p_i ({\bf x} \cdot {\bf p})
\label{eqn:conqs:H}
\ea
are constants of motion for the system $(N, {\tilde {\omega}}, H)$, i.e.,
${\bf {\hat X}}_{H}(J_i)={\bf {\hat X}}_{H}(A_i)=0$.
The quantities $A_i$ are quadratic in the momenta and so do not arise as a
result of any simple Killing vector symmetry of the manifold ${\cal E}^3$.

Finally, we note that the Poisson brackets of the constants of motion $J_i$ and $A_i$ are
\ba
\{ J_i , J_j \} &=& -\epsilon_{ij}^k J_{k}, \qquad
\{ J_i , A_j \} = -\epsilon_{ij}^k A_{k}, \nonumber\\
\{ A_i , A_j \} &=& H \, \epsilon_{ij}^k \; J_{k}.
\label{eq:liealg}
\ea
Therefore, on constant energy hypersurfaces $H=E$ the Lie algebra given by the Poisson brackets is isomorphic to $so(4)$ for the case
$E<0$, $so(3,1)$ for $E >0$, and $iso(3)$ for $E=0$.
Thus there is a one-to-one correspondence between the dynamical symmetry
algebras of motion in the classical Kepler problem and geodesic motion on
three-dimensional spaces of constant curvature.

\section{Geodesic motion on spaces of constant curvature}
\label{sec:geomotconcurgeo}
We shall now present the Hamiltonian function, phase flow and constants of
motion for geodesic motion on spaces of constant curvature and indicate how
they are related to those corresponding to classical Kepler motion.
We can write the line element for a three-dimensional space of constant curvature
$k$ in terms of the stereographic coordinates $\{ {x}^i \}, i=1,2,3$ as
\be
 ds^2 =  \biggl(1+ {{k|{\bf x}|^2} \over 4} \biggr)^{-2} \: ds^2_{\cal E}
\label{eq:sccmetric}
\ee
where $ds^2_{\cal E}=\delta_{ij} d{x}^i d{x}^j$ and
$|{\bf {x}}|^2 = \delta_{ij} {x}^i {x}^j$.
We denote the generic three-geometry as ${\cal G}^3(k)$.

Geodesic motion on such a space can be described by a Hamiltonian
\be
G = {1 \over 2} \biggl(1+{{k|{\bf {x}}|^2} \over 4} \biggr)^2 \,
|{\bf {p}}|^2
\label{eqn:hamfunc:scc}
\ee
where the three-vectors ${\bf {x}}$, ${\bf {p}}$ represent the position and velocity vectors, respectively, of the particle.
Thus we label this Hamiltonian system
$({\bar N}, {\bar \omega}, G)$ where
${\bf {\bar \omega}} = {\bf d} p_i \wedge {\bf d} x^i $.

The homogeneity and isotropy of these three-dimensional spaces give rise to a
six-dimensional group of isometries and the corresponding Killing vectors form a six-dimensional Lie algebra. The subgroup $SO(3)$ is common to all three types
of geometry ${\cal S}^3$, ${\cal E}^3$ and ${\cal H}^3$.
A basis for the associated Lie algebra of Killing vectors is
\be
{\bf R}_i =  \epsilon_{ij}^k \, {{x}^j}{\partial  \over{\partial {x}^k}}.
\ee
The homogeneity is granted through invariance of the metric
(\ref{eq:sccmetric}) under transitive motions. However, the transitive
subgroup is different according to the value of $k$. A basis for the associated
Lie algebra of Killing vector fields is
\be
{\bf P}_i = \biggl(1-{{k|{\bf {x}}|^2} \over 4} \biggr)
{\partial \over {\partial {{x}^i}}} \: + \: {k \over 2} \; {x}^i \,
\biggl({x}^j {\partial \over {\partial {{x}^j}}}\biggr).
\ee
The associated constants of motion $L_i={\bf R}_i \cdot {\bf p}$ and
$D_i={\bf P}_i \cdot {\bf p}$ are respectively
\ba
{L_i} & = &  \epsilon_{ij}^k \, {{x}^j}{p}_k ,
\label{eqn:Li} \\
{D_i} & = & \biggl(1-{{k|{\bf {x}}|^2} \over 4} \biggr) \; {p}_i
+ {k \over 2} \, x^i\; ({\bf {x}} \cdot {\bf {p}}),
\label{eqn:Di}
\ea
satisfying
${\bf {\hat X}}_{G}(L_i)={\bf {\hat X}}_{G}(D_i)=0$.
The constants of motion have Poisson brackets with structure constants
identical to those of the Killing vectors and are as follows
\ba
  \{ L_i,L_j \} & = & {-{\epsilon_{ij}^k} {L_k}},  \qquad
  \{ L_i,D_j \} = {-{\epsilon_{ij}^k} {D_k}},   \nonumber\\
  \{ D_i,D_j \} & = & {-k{\epsilon_{ij}^k} {L_k}}.
\label{eq:frwconalg}
\ea
Thus, the functions $L_i,D_j$ form a Lie algebra under the Poisson bracket
operation depending on the value of $k$.
It can be seen that the Lie algebra given by the Poisson brackets is isomorphic
to the Lie algebra $so(4)$ for the case $k>0$, $so(3,1)$ for $k<0$, and $iso(3)$ for $k=0$.

We now implement a canonical transformation on the phase space as
described in the Appendix.
The Hamiltonian function becomes
\be
G= {1 \over 4} \biggl(k+{|{\bf {\bar p}}|^2 \over 2} \biggr)^2
|{\bf {\bar x}}|^2
\label{eqn:hamsccII}
\ee
and the Hamiltonian vector field corresponding to the phase flow is
\be
{\bf {\hat X}}_{G} =
 G^{1 \over 2} \, |{\bf {\bar x}}| \, \biggl( {\bar p}_i
{{\partial \phantom{z}} \over {\partial {\bar x}^i}}
-
2 \, G^{1 \over 2}{{\bar x}^j \over \phantom{|}|{\bf {\bar x}}|^3} \;
{{\partial \phantom{z}} \over {\partial {\bar p}_j}} \biggr).
\label{eqn:hamvecscc}
\ee
We note that ${\bf {\hat X}}_{G}$ has a similar form to that for
Kepler motion (\ref{eqn:hamvecckp}), except for the factor
$G^{1 \over 2} \, |{\bf {\bar x}}|$. The constants of motion are
\ba
{L_i} =  \epsilon_{ij}^k \, {\bar x}^j {\bar p}_k , \nonumber\\
\sqrt{2}D_i = {\bar x}^i
\biggl(|{\bf {\bar p}}|^2 - {2G^{1 \over 2} \over {|{\bf {\bar x}}|}} \biggr)
-  {\bar p}_i ({\bf {\bar x}} \cdot {\bf {\bar p}}) ,
\ea
which are similar to those associated with Kepler motion presented
in (\ref{eqn:conqs:H}).
At this point we note that (\ref{eqn:hamsccII}) can easily be rearranged to
give
\be
-k={{|{\bf {\bar p}}|} \over 2}^2-{{2G^{1 \over 2}} \over
{|{\bf {\bar x}}|}}
\label{eqn:hamkeppro}
\ee
which resembles the Kepler problem Hamiltonian function. Expressions
(\ref{eqn:hamsccII}) - (\ref{eqn:hamkeppro}) form the basis of the mapping
discussed in section \ref{sec:dynsys}.

\section{Related dynamical systems}\label{sec:dynsys}
Consider for the moment the constant energy surface $G=C$ in system
$({\bar N}, {\bar {\omega}}, G)$.
If we identify the constants $\alpha=2 C^{1 \over 2}$ then the
Hamiltonian vector fields (\ref{eqn:hamvecckp}) and (\ref{eqn:hamvecscc})
have the same form, apart from a factor $C^{1 \over 2} \, |{\bf {\bar x}}|$.
Thus if we can identify the phase space coordinates under some map then,
for the constant energy surface $H=E$ in the Hamiltonian system
$(N, {\tilde {\omega}}, H)$, we can say that $E=-k$ and we can see that the Hamiltonian vector fields can be considered parallel.
We point out that the case where $\alpha=2 C^{1 \over 2}=0$ must be excluded from this analysis.
If we write
${\bf {\hat X}}_{G} = {d / d\lambda}$ and ${\bf {\hat X}}_{H} = {d / d\tau}$,
then the relationship between the Hamiltonian vector fields
(\ref{eqn:hamvecckp}) and (\ref{eqn:hamvecscc}) can be written
\be
{d \over d\lambda} =  C^{1 \over 2} \, |{\bf {\bar x}}| \,
{d \over d\tau}
\label{eqn:t+tau}
\ee
and so the time parameters are related by the following
${d\tau / d\lambda} \: = \: C^{1 \over 2} \, |{\bf {\bar x}}|$.
This parameter change is part of Moser's transformation, see equation (2.8)
in \cite{moser70}.
It can also be seen that upon this identification, the quantities $A_i$ and
$D_i$ have precisely the same form. It is clear that the quantities $J_i$ and
$L_i$ have the same form.

Thus we can see from (\ref{eqn:hamkeppro}) that such a mapping will
result in the energy $E$ in the Kepler problem being given by $E=-k$, i.e.,
the whole of phase space for motion on a space with curvature $k$ will be
mapped to energy surfaces for Kepler problems with different mass
parameters $\alpha$ but with {\it the same} energy $E$.
It then follows from this that energy surfaces $G=C$ for motion on a space of
fixed $k$ will be mapped to energy surfaces $E=-k$ for Kepler motion with
fixed mass parameter $\alpha=2C^{1 \over 2}$, i.e., energy surfaces are mapped
to energy surfaces as in Moser \cite{moser70} and Belbruno \cite{belbruno77}.

We shall now formulate these notions rigourously and consider the implications
of this result. First we shall construct a map $\psi$ between the
$2n$-dimensional phase spaces $N$ and ${\bar N}$
\be
\psi : {\bar N} \mapsto N
\label{eqn:mappsi}
\ee
such that $x^i={\bar x}^i$ and $p_j={\bar p}_j$. Thus we have that
${\bf d}p_i \wedge {\bf d}x^i = {\bf d}{\bar p}_j \wedge {\bf d}{\bar x}^j$.
Thus the manifolds $N$ and ${\bar N}$ have the same symplectic structure.

Now we can construct the $2n+1$ dimensional evolution space
$W^\alpha=\mathbf{R} \times N$ with closed two-form
\be
{\bf \Omega}={\bf d}p_i \wedge {\bf d}x^i - {\bf d}H \wedge d\tau,
\ee
where $\tau$ is the time parameter for the system. This
evolution space is characterised by the parameter $\alpha$ via the Hamiltonian function $H$. Thus we can construct the family of evolution spaces, i.e., the
fibre bundle ${\cal F}= \mathbf{R} \times W^\alpha$. This fibre bundle can be
regarded as {\it the space of all Kepler problems}, each characterised by
the value of $\alpha$.
Similarly we can construct the evolution space
${\bar W}^k=\mathbf{R} \times {\bar N}$ with closed two-form
\be
{\bf {\bar \Omega}}={\bf d}{\bar p}_i \wedge {\bf d}{\bar x}^i -
{\bf d}G \wedge d\lambda
\ee
characterised by the constant $k$. The corresponding family of evolution spaces is ${\bar {\cal F}}= \mathbf{R} \times {\bar W}^k$. This fibre bundle can be
regarded as {\it the space of all systems of geodesic motion on spaces of
constant curvature}.

Now, ${\bar {\cal F}}|^k={\bar W}^k$ is a hypersurface in ${\bar {\cal F}}$
corresponding to a particular value of $k$. We can define an energy
hypersurface via the map
${\bar \phi}: {\bar {\cal F}}|^k_C \mapsto {\bar {\cal F}}|^k$
to be the surface defined by $G=C$, i.e.,
\be
-k=|{\bf {\bar p}}|^2/2-2C^{1 \over 2}/|{\bf {\bar x}}|.
\ee
Similarly, ${{\cal F}}|^\alpha=W^\alpha$ is a hypersurface in the
fibre bundle ${\cal F}$ and we can define the energy surface via the map
$\phi: {\cal F}|^\alpha_E \mapsto {\cal F}|^\alpha$ to be the surface $H=E$,
i.e.,
\be
E=|{\bf p}|^2/2-\alpha/|{\bf x}|.
\ee
Alternatively, we can define a hypersurface in the fibre bundle ${\cal F}$
corresponding to the space of all orbits with energy $H=E$ via a map
$\pi: {\cal F}|_E \mapsto {\cal F}|$.

We then define a map $\Phi$ between energy hypersurfaces in the
respective evolution spaces as follows
\be
\Phi : {\bar {\cal F}}|^{k} \mapsto {\cal F}|_E
\label{eqn:map1}
\ee
such that $E=-k$. Thus under the map $(\Phi \circ \psi)$ we find that
$\alpha=2C^{1 \over 2}$ and that the surface $ {\bar {\cal F}}|^k_C$
is mapped into the surface ${\cal F}|^\alpha_E$.
Under the the map $\psi$ (\ref{eqn:mappsi})
we find that ${\bf \tilde{\omega}} = {\bf {\bar {\omega}}}$ and so
$(\Phi \circ \psi)^*{\bf {\bar {\omega}}}={\bf \tilde{\omega}}$ and under the
map $\Phi$ we have as a result of (\ref{eqn:t+tau}) that
${\bf d}H \wedge {\bf d}\tau={\bf d}G \wedge {\bf d}\lambda$ and so
\be
(\Phi \circ \psi)^* \, {\bar \Omega} = \Omega .
\label{eqn:mapomega}
\ee
Now consider the surface ${\bar {\cal F}}|^{k}_{C}$ in ${\bar {\cal F}}$.
We have established that the quantities $L_i$ and $D_j$ have Poisson brackets
given by (\ref{eq:frwconalg}) and that the symplectic two-form
${\bf {\bar {\omega}}}$ is preserved under the map $(\Phi \circ \psi)$.
Since the form of the quantities $L_i$ and $D_j$ are preserved under the map
$(\Phi \circ \psi)$, these quantities have the same poisson brackets
on the surface ${\cal F}|^\alpha_E$ in ${\cal F}$.
Under the map we can write
\be
(\Phi \circ \psi)_* \, {\bf {\hat X}}_{H} = {1 \over C^{1 \over 2} \,
|{\bf {\bar x}}|}{\bf {\hat X}}_{G}
\label{eqn:xgfxh}
\ee
and so the quantities $L_i$ and $D_j$ will also be constants of the motion
on the energy surface $H=E=-k$, $\alpha = 2C^{1 \over 2}$ in the
system $(N, {\tilde {\omega}}, H)$. Since the poisson structure is
preserved then their Poisson bracket Lie algebra will be identical to that
in (\ref{eq:frwconalg}). It is indeed the case that the quantities $J_i=L_i$
and $A_i=D_i$ under the map $(\Phi \circ \psi)$.

Thus, given any point (${\bar x}^i,{\bar p}_j,\lambda,k$) in the fibre bundle
${\bar {\cal F}}$, we have defined the corresponding point
(${x}^i,{ p}_j,\tau,\alpha$) in the Kepler problem fibre bundle ${\cal F}$ by
$x^i={\bar x}^i$, $p_j={\bar p}_j$, $\alpha=+2C^{1 \over 2}$ and
$\tau=\tau(\lambda,x^i,p_j)$ from relation (\ref{eqn:t+tau}).
This mapping is injective. The cases $\alpha > 0$
and $\alpha < 0$ must be treated separately and in the latter case
we would define a map with $\alpha=-2C^{1 \over 2}$. In a sense, the Kepler
problem is the ``square root" of geodesic motion on spaces of constant
curvature in the same way as spinors are the ``square root" of tensors. We note
that this suggests considering the Kepler problem in a complex space.
Two-body motion with a central repelling field, that is, the case where
$\alpha < 0$, necessarily implies that $H > 0$ and so $k < 0$ which corresponds
to the hyperbolic geometry ${\cal H}^3$.

Thus the phase flows in the surfaces ${\bar {\cal F}}|^{k}_{C}$
are mapped into those in ${\cal F}|^\alpha_E$ where $E=-k$ and
$\alpha=+2C^{1 \over 2}$ (or $\alpha=-2C^{1 \over 2}$) and so the constants of
motion in the first will be mapped into constants of the motion in the second.
The formalism presented above explains why the Poisson bracket Lie algebra of
constants of motion on constant energy surfaces in the classical Kepler problem
is identical to that for the constants for geodesic motion on spaces of
constant curvature.
In addition it is clear that replacing the constant $k$ by the function
$-H$ in the constants of motion $D_i$ gives the components of the Runge-Lenz
vector $A_i$.
Hence, the constants of motion $A_i$ in the classical Kepler problem arise in a
natural way from the transitive isometries of the associated spaces of
constant curvature. Furthermore, the commutation relations amongst the
constants of motion in the Kepler problem
$(N, {\bf \tilde{\omega}}, H)$ can be thought of as being inherited
from those in $({\bar N}, {\bar {\omega}}, G)$, i.e., the commutation relations
(\ref{eq:frwconalg}), when restricted to the appropriate energy surface.

So far we have excluded the singularity at the origin $|{\bf x}|=0$ in the Kepler problem.
This is equivalent to exclusion of certain points in the
corresponding spaces ${\cal G}^3(k)$.
It is now straightforward to reinstate these points thereby regularising the problem. For example, the $E < 0$ energy surface in
the Kepler problem is mapped to the tangent bundle of the sphere ${\cal S}^3$
punctured at one point, the north pole. The $E < 0$ Kepler problem energy
surface is compactified when we include this point and so the geodesics through
the north pole are transformed into collision orbits.

\section{Conclusions}
The formalism we have presented generalises and we hope, clarifies previous work. The main difference is that there are two parameters in each problem:
the mass parameter $\alpha$ and the value of the Hamiltonian $H$ in the Kepler
problem and the curvature $k$ and the value of the Hamiltonian $G$ in the
system of geodesic motion on spaces of constant curvature.
In this paper we have constructed the family of evolution spaces corresponding to both
problems and via a suitable map shown the equivalence of the flows in the two
problems [equations (\ref{eqn:mapomega}) and (\ref{eqn:xgfxh})].
Within this framework the energy of the Kepler problem is equal to (minus)
the curvature parameter of the constant curvature space, i.e., $H=-k$, and the mass parameter
is given by the value of the Hamiltonian for the geodesic motion problem via the relation
$\alpha=2G^{1 \over 2}$.
This ensures the correspondence of the constants of motion for both systems.
This result gives the Runge-Lenz vector a ``purely geometrical"
interpretation in that it arises from isometries of a three-dimensional manifold
associated with the Kepler Hamiltonian system.
In addition, the canonical nature of the transformation guarantees the correspondence between the
Poisson bracket Lie algebras of constants of motion on the constant energy hypersurfaces
of the two systems.
We have presented a unified treatment which is valid for all values of energy
continuously. This is in contrast to the work of Moser \cite{moser70} and Belbruno
\cite{belbruno77} which dealt with a {\it single} Kepler problem and necessitated the
use of a different procedure for each sign of energy.
Our method highlights the relationship between the two phase flows in a
{\it geometrical} way and we have used clear and simple canonical
transformations to obtain the result.
Finally we note that we have said only that the two systems are related and we have
{\it not} transformed one system into the other. In particular, $H$ and $G$ are {\it not} the same Hamiltonian.

It is apparent from this result that the curvature parameter $k$ may be considered
to be a coordinate in some higher dimensional space and that this may provide further
information about the dynamics of the classical Kepler problem. In addition, it is well
known that the spectrum generating Lie algebra and Lie algebra corresponding to the dynamical
group of the classical Kepler problem is the Lie algebra $so(4,2)$. This Lie algebra also
appears as the conformal symmetry group of Minkowski spacetime and our formalism
suggests a connection between the two. These issues are
currently being investigated by the authors.

This type of procedure may have applications to other dynamical systems with
dynamical symmetries, for example, the harmonic oscillator and systems which admit
Killing tensors and associated quadratic constants of motion.
That is, there may be a geometrical interpretation available for the dynamical constants of motion which exist in these systems.

\appendix
\section*{Appendix: Canonical transformation on $T^*{\cal G}^3(k)$}
\setcounter{section}{1}
\label{sec:cantransf}
It turns out that for our purposes it is simplest to transform from the
natural coordinate system of (\ref{eq:sccmetric}) to a new coordinate system
on configuration space.
We perform an inversion of the coordinates
\be
x'^i= { {{x}^i} \over {|{\bf {x}}|^2}}.
\ee
The corresponding canonical momenta
($p'_i= {{\partial {x}^j} / {\partial {{x'}^i}}} \, {p}_j$) are
\be
p'_i = |{\bf {x}}|^2 {{p}_i}-2{{x}^i}({\bf {x}} \cdot {\bf {p}}).
\ee
Since this is just a coordinate transformation on configuration space, it
is obviously canonical.
Using these inverted coordinates makes the relationship between the
classical Kepler motion and geodesic motion on surfaces of constant curvature
much more transparent. For flat Euclidean space ${\cal E}^3$, i.e., the case
$k=0$, this coordinate inversion takes points near the origin to points near infinity and
vice versa, but of course the intrinsic geometry of the space is
unaltered. A similar argument holds for the case ${\cal H}^3$, $k < 0$ as
shown by Belbruno \cite{belbruno77}.
For the three-sphere ${\cal S}^3$ with $k > 0$ this corresponds to an antipodal
mapping which is an isometry of ${\cal S}^3$. This is why Moser's method \cite{moser70} does
not require the inversion.
Then we implement a further (canonical) transformation
\be
{\bar x}^i = p'_i / 2\sqrt{2} , \qquad
{\bar p}_i=-2\sqrt{2} {x'}^i.
\ee
This transformation relates the momentum space of the classical Kepler
problem to a configuration space of constant curvature, that is, the
velocity hodographs can be regarded as geodesics on such spaces,
as discussed in Moser \cite{moser70} and Belbruno \cite{belbruno77}.

Combining these two transformations the Hamiltonian function becomes
\be
G= {1 \over 4} \biggl(k+{|{\bf {\bar p}}|^2 \over 2} \biggr)^2
|{\bf {\bar x}}|^2
\ee
and the conserved quantities are
\ba
{L_i} = \epsilon_{ij}^k \, {\bar x}^j {\bar p}_k , \\
\sqrt{2}D_i = {\bar x}^i
\biggl(|{\bf {\bar p}}|^2 - {2G^{1 \over 2} \over {|{\bf {\bar x}}|}} \biggr)
-  {\bar p}_i ({\bf {\bar x}} \cdot {\bf {\bar p}}).
\ea

\end{document}